\documentclass[aps,prl,twocolumn,superscriptaddress]{revtex4}
\usepackage{graphicx}
\usepackage{mathrsfs}
\usepackage{bm}
\usepackage{amsmath}
\usepackage{dcolumn}
\usepackage{epstopdf}
\usepackage{dsfont}
\usepackage{amssymb}
\usepackage{tabularx}
\usepackage{array}
\usepackage{colordvi}
\usepackage{txfonts}

\begin{document}
\title{Universal Current Correlations Induced by the Majorana and Fermionic Andreev Bound States}
\author{Kunhua Zhang}
\affiliation{ICQD, Hefei National Laboratory for Physical Sciences at Microscale, and Synergetic Innovation Center of Quantum Information and Quantum Physics, University of Science and Technology of China, Hefei, Anhui 230026, China.}
\affiliation{CAS Key Laboratory of Strongly-Coupled Quantum Matter Physics, and Department of Physics, University of Science and Technology of China, Hefei, Anhui 230026, China.}

\author{Xinlong Dong}
\affiliation{School of Chemistry and Materials Science, Shanxi Normal University, Linfen, Shanxi 041004, China.}
\affiliation{ICQD, Hefei National Laboratory for Physical Sciences at Microscale, and Synergetic Innovation Center of Quantum Information and Quantum Physics, University of Science and Technology of China, Hefei, Anhui 230026, China.}

\author{Junjie Zeng}
\affiliation{ICQD, Hefei National Laboratory for Physical Sciences at Microscale, and Synergetic Innovation Center of Quantum Information and Quantum Physics, University of Science and Technology of China, Hefei, Anhui 230026, China.}
\affiliation{CAS Key Laboratory of Strongly-Coupled Quantum Matter Physics, and Department of Physics, University of Science and Technology of China, Hefei, Anhui 230026, China.}

\author{Yulei Han}
\affiliation{ICQD, Hefei National Laboratory for Physical Sciences at Microscale, and Synergetic Innovation Center of Quantum Information and Quantum Physics, University of Science and Technology of China, Hefei, Anhui 230026, China.}
\affiliation{CAS Key Laboratory of Strongly-Coupled Quantum Matter Physics, and Department of Physics, University of Science and Technology of China, Hefei, Anhui 230026, China.}

\author{Zhenhua Qiao}
\email[Correspondence to:~~]{qiao@ustc.edu.cn}
\affiliation{ICQD, Hefei National Laboratory for Physical Sciences at Microscale, and Synergetic Innovation Center of Quantum Information and Quantum Physics, University of Science and Technology of China, Hefei, Anhui 230026, China.}
\affiliation{CAS Key Laboratory of Strongly-Coupled Quantum Matter Physics, and Department of Physics, University of Science and Technology of China, Hefei, Anhui 230026, China.}

\begin{abstract}
Due to their nonlocality, Majorana bound states have been proposed to induce current-current correlations (CCCs) that are completely different from those induced by low-energy fermionic Andreev bound states. Such characteristics can be used as a signature to detect Majorana bound states. Herein, we studied the Majorana and fermionic Andreev bound states in a two-dimensional topological insulator system. We found that nonlocality occurs for both types of bound states and that their coupling strengths depend on system parameters in the same pattern. Majorana and fermionic Andreev bound states show the same differential CCCs characteristics, thereby indicating a universal behavior for both types of bound states. The maximal cross differential CCCs are robust to the structural asymmetry of the system.
\end{abstract}

\maketitle

\emph{Introduction.---}In condensed matter systems, Majorana bound states (MBSs) are exotic excitations of zero energy. They are their own antiparticles because of the equal superposition of the electron and hole excitations~\cite{F.Wilczek2009,S.Elliott2015}. Two well-separated MBSs store information nonlocally, making the information immune to local perturbations~\cite{A.Kitaev2001}. In addition to complying with non-Abelian statistics~\cite{D.Ivanov2001,C.Nayak2008}, MBSs have potential applications in decoherence-free quantum computation~\cite{A.Kitaev2003}. Among various condensed matter systems, topological superconductors represent a natural means of searching MBSs and therefore have recently attracted considerable attention~\cite{G.Moore1991,S.Sarma2006,S.Tewari2007}. Proposals have been made to realize topological superconductors in a variety of candidate systems wherein superconductivity is obtained as a result of the proximity effect of an $s$-wave superconductor~\cite{L.Fu2008,L.Fu2009a,J.Sau2010,J.Alicea2010,R.Lutchyn2010,Y.Oreg2010,N.Perge2012}. Multiple studies have been conducted to verify the existence of MBSs in various topological superconductor systems~\cite{K.Sengupta2001,K.T.Law2009,A.R.Akhererov2009,L.Fu2009b,K.Flensberg2010,S.Sarma2012,J.J.He2014a,J.J.He2014b,X.Liu2015}. Through experiments, some evidence has been found for the existence of MBSs owing to phenomena such as resonant Andreev reflection, fractional Josephson effect, selective equal-spin Andreev reflection, and half-integer conductance plateau~\cite{V.Mourik2012,M.Deng2012,L.P.Rokhinson2012,H.H.Sun2016,N.Perge2016,Q.L.He2017,H.Zhang2018}. However, because these phenomena have possible physical explanations, except for MBSs, more compelling experimental evidence regarding these signatures is required to settle the debate on MBSs~\cite{J.Liu2012,D.Pikulin2012,E.Prada2012,E.Lee2012,W.Ji2018,Y.H.Li2018}.

As a unique property of MBSs, their nonlocality gives rise to nonlocal transport if there is coupling among them. Such coupling comprises Coulomb coupling and tunneling coupling, which exist due to the charging energy and the overlap of wave functions, respectively~\cite{C.Bolech2007,J.Nilsson2008,L.Fu2010,Beenakker2013a,S.M.Albrecht2016}. Herein, we study crossed Andreev reflection for the case of tunneling coupling~\cite{A.F.Andreev1964,J.M.Byers1995,A.R.Akhererov2009}. When MBSs are strongly coupled, local Andreev reflection is predicted to be completely suppressed at sufficiently low excitation energy while favoring crossed Andreev reflection. A characteristic of this enhanced crossed Andreev reflection is maximal cross current-current correlation (CCC)~\cite{J.Nilsson2008}.

Moreover, other studies have shown that the CCCs induced by MBSs differ from those induced by ordinary low-energy fermionic Andreev bound states (ABSs)~\cite{J.Liu2013,Y.M.Wu2015}. Therefore, a question arises as to whether the enhanced crossed Andreev reflection and maximal cross CCC are unique to MBSs. To answer this question, we construct MBSs and fermionic ABSs in a two-dimensional topological insulator (2D TI) system. We find that nonlocality occurs for both MBSs and ordinary fermionic ABSs and that the coupling strengths of these bound states depend on system parameters displaying the same tendency. When the integral effect on the bias voltage is removed, these bound states result in the same maximal cross differential CCCs. Such correlations are universal.
\begin{figure}[t]
\includegraphics[width=0.45\textwidth]{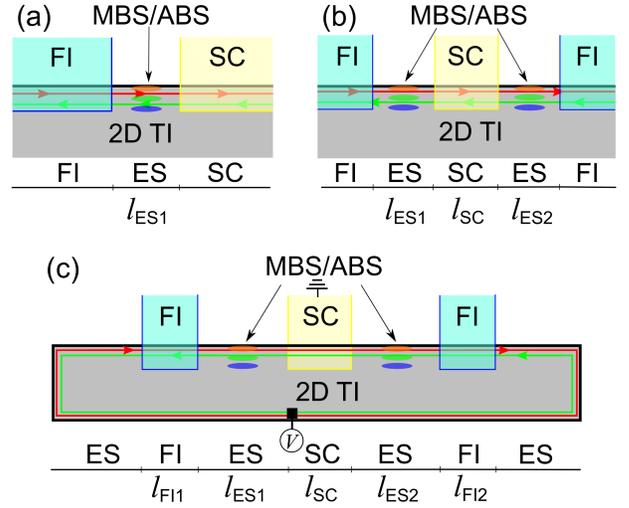}
\caption{(Color online) Schematics of one-dimensional ferromagnetic-insulator--edge-state--superconductor (FI--ES--SC) junctions mediated by the ESs of a two-dimensional topological insulator (2D TI): (a) FI--ES--SC junction; (b) FI--ES--SC--ES--FI junction; (c) ES--FI--ES--SC--ES--FI--ES junction.}
\label{fig1}
\end{figure}

\emph{Model.---}We consider MBSs and ordinary fermionic ABSs in one-dimensional ferromagnetic-insulator--edge-state--superconductor (FI--ES--SC) junction systems mediated on the edge of a 2D TI, as shown in Fig.~\ref{fig1}. The ferromagnetism and superconductivity of the ESs are induced by the proximity effects of the FI and the $s$-wave SC, respectively, which interact with the electrons in the ESs of a 2D TI~\cite{L.Fu2009a,I.Knez2012}.

The one-dimensional junctions can be described by the following Bogoliubov--de Gennes equation~\cite{L.Fu2009a,J.Nilsson2008}:
\begin{align}
\label{BdG Hamiltonian}
\begin{pmatrix}
 \upsilon_{\rm F}\sigma_{x} p_{x}+\bm{\sigma}\cdot\bm{m}-\mu&\Delta \mathrm{e}^{\mathrm{i}\phi}\\
 \Delta \mathrm{e}^{-\mathrm{i}\phi}&-\upsilon_{\rm F}\sigma_{x} p_{x}+\bm{\sigma}\cdot \bm{m}+\mu
 \end{pmatrix} \psi = E\psi,
\end{align}
where $\bm{\sigma} = (\sigma_{x},\sigma_{y},\sigma_{z})$, $\upsilon_{\rm F}$, $\psi$, and $E$ are the Pauli matrices, Fermi velocity, wave function, and excitation energy, respectively. $\mu(x)$ is the chemical potential measured with respect to the Dirac point. $\Delta \mathrm{e}^{\mathrm{i}\phi}$ denotes the superconducting pair potential, where $\Delta$ and $\phi$ are the energy gap and the phase, respectively. Because $\phi$ makes no difference to the calculations, we set it to be zero. In Fig.~\ref{fig1}(a), the magnetization is $\bm{m}(x) = (m_{lx},m_{ly},m_{lz})$ for $ x < -l_{\rm{ES1}}$ and $\bm{m}(x) = 0$ otherwise. In Fig.~\ref{fig1}(b), the magnetization is set as $\bm{m}(x) = (m_{lx},m_{ly},m_{lz})$ for $ x < -l_{\rm{ES1}}$ and $\bm{m}(x) = (m_{rx},m_{ry},m_{rz})$ for $x > l_{\rm{SC}}+l_{\rm{ES2}}$.

The Fermi level is uniform in the whole junction, but the chemical potential $\mu$ is position dependent and can be tuned by the gate voltage or doping in each region~\cite{I.Knez2012}. In the following, $\mu_{\rm{EG}1}$ and $\mu_{\rm{EG}2}$ denote the chemical potentials at the left and right ESs around the SC, respectively. The chemical potentials for the SC and the left and right FIs are represented by $\mu_{\rm{SC}}$, $\mu_{\rm{FI}1}$, and $\mu_{\rm{FI}2}$, respectively.

By solving Eq.~(\ref{BdG Hamiltonian}), we obtain the wave functions for the junctions shown in Fig.~\ref{fig1}. For example, the wave function of the FI in Fig.~\ref{fig1}(a) can be expressed as follows:
\begin{equation}
\psi_{\rm{FI}1}=a_{e}\psi_{\text{FI}1}^{e} \exp[{-\mathrm{i}(k_{l}+\frac{2m_{lx}}{\hbar\upsilon_{\text{F}}})x}]+a_{h}\psi_{\text{FI}1}^{h} \exp[{\mathrm{i} (\frac{2m_{lx}}{\hbar\upsilon_{\text{F}}}-k^{\prime}_{l})x}],
\end{equation}
where $\psi_{\text{FI}1}^{e} = (-\hbar\upsilon_{\rm F}k_{l}-m_{lx}-\mathrm{i} m_{ly},E+\mu_{\rm{FI}1}-m_{lz},0,0)^{T}$ and $\psi_{\text{FI}1}^{h} = (0,0,\hbar\upsilon_{\rm F}k^{\prime}_{l}-m_{\rm{lx}}-\mathrm{i} m_{\rm{ly}}, E-\mu_{\rm{FI}1}-m_{lz})^{T}$. $T$ indicates matrix transposition. We set the chemical potential $\mu_{\rm{FI}1}$ to zero at the Dirac point. Then, $k_{l} = \left(\mathrm{i}\sqrt{-(E+\mu_{\rm{FI}1})^2+m_{lz}^2+m_{ly}^2}-m_{lx}\right)/\hbar\upsilon_{\text{F}}$ and
$k^{\prime}_{l} = \left(\mathrm{i}\sqrt{-(E-\mu_{\rm{FI}1})^2+m_{lz}^2+m_{ly}^2}+m_{lx}\right)/\hbar\upsilon_{\text{F}}$. $a_{e}$ and $a_{h}$ are the coefficients of the electron and hole wave functions, respectively.

The ES wave function is expressed as follows:
\begin{eqnarray}
\psi_{\rm{EG}1} &=& b_{e} \psi_{\text{ES}1}^{e}\exp(\mathrm{i}k_{1}x)+b_{e}^{\prime} \psi_{\text{ES}1}^{e\prime}\exp(-\mathrm{i}k_{1}x)\nonumber\\
&+&c_{h} \psi_{\text{ES}1}^{h}\exp(\mathrm{i}k_{2}^{\prime}x)+c_{h}^{\prime} \psi_{\text{ES}1}^{h\prime}\exp(-\mathrm{i}k_{2}x),
\end{eqnarray}
where $b_{e}$, $b_{e}^{\prime}$, $c_{h}$, and $c_{h}^{\prime}$ are the coefficients of wave functions. $\psi_{\text{ES}1}^{e} = (\hbar\upsilon_{\text{F}}k_{1},E+\mu_{\rm{EG}1},0,0)^{T}$, $\psi_{\text{ES}1}^{e\prime} = (-\hbar\upsilon_{\text{F}}k_{1},E+\mu_{\rm{EG}1},0,0)^{T}$, $\psi^{\rm{h}}_{t} = (0,0,-\hbar\upsilon_{\text{F}}k_{2},
E-\mu_{\rm{EG}1})^{T}$, and $\psi^{\rm{h\prime}}_{t} = (0,0,\hbar\upsilon_{\text{F}}k_{2}^{\prime},
E-\mu_{\rm{EG}1})^{T}$. Here $k_{1} = \mu_{\rm{EG}1}+E$, and $k_{2} = \mu_{\rm{EG}1}-E$.

The SC wave function is expressed as follows:
\begin{equation}
\psi_{\rm{SC}} = d \psi_{\text{SC}}^{1} \exp[{(-\kappa-\mathrm{i}k_{\text{SC}})x}]+f\psi_{\text{SC}}^{2} \exp[{(-\kappa+\mathrm{i}k_{\text{SC}})x}],
\end{equation}
where $\psi_{\text{SC}}^{1,2} = (\mp \exp[\mathrm{i}(\phi \mp \alpha)],\exp[\mathrm{i}(\phi\mp\alpha)],\mp1,1)^{T}$, $k_{\text{SC}} = \mu_{\rm SC}/\hbar\upsilon_{\rm F}$, $\alpha = \arccos(E/\Delta)$ for $E<\Delta$, and $\kappa = \Delta\sin\alpha/\hbar\upsilon_{\text{F}}$. $d$ and $f$ are coefficients of wave functions that are coherent superpositions of the electron and hole excitations. The wave functions in different regions satisfy continuity at the interfaces, which determines the properties of the bound states. With the same method, we can obtain the wave functions and the properties of the bound states shown in Fig.~\ref{fig1}(b).

\begin{figure}[t]
\includegraphics[width=0.47\textwidth]{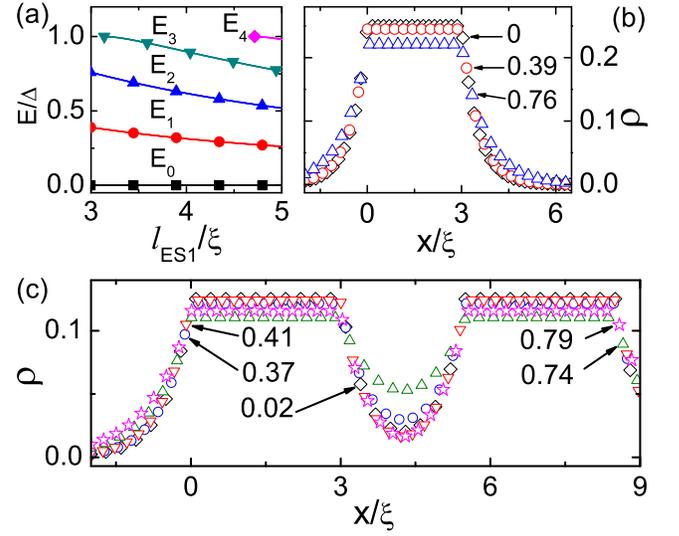}
\caption{(Color online) (a) Energies $E$ of the bound states as functions of the ES width of the FI--ES--SC junction $l_{\rm{ES}1} = 3\xi$. (b) Probability densities $\rho$ of Majorana bound states (MBSs) and ordinary Andreev bound states (ABSs) as functions of $x$, with $x = 0$ as the interface between the leftmost FI and ESs. Here, the chemical potential $\mu_{\rm{SC}} = 50\Delta$. In (b) and (c), $l_{\rm{SC}} = 2.5\xi$ and $l_{\rm{ES}1} = l_{\rm{ES}2} = 3\xi$. $\xi = \hbar\upsilon_{\rm F}/\Delta$ is the coherence length.}
\label{fig2}
\end{figure}
\emph{MBSs and fermionic ABSs.---}First, we study the MBSs and non-zero-energy fermionic ABSs in the junction shown in Fig.~\ref{fig1}(a). Because the electron spin is locked with the momentum of ESs in the 2D TI, both the magnetization and $s$-wave superconducting pair potential can open gaps in the gapless ESs. As shown in Fig.~\ref{fig1}(a), if the FI and SC are infinitely long, bound states can exist in this junction. Based on the wave functions and boundary conditions, the energies $E$ and probability densities $\rho$ of all bound states can be calculated. As shown in Fig. 2(a), the number of bound states increases discontinuously with an increase in the width $l_{\rm{ES}1}$ of the junction. The zero-energy bound states (i.e., MBS) always exists and is independent of $l_{\rm{ES}1}$, whereas the energies of the non-zero-energy bound states (i.e., ordinary fermionic ABSs) decrease with an increase in $l_{\rm{ES}1}$.

Figure~\ref{fig2}(b) shows the probability densities $\rho$ of the three bound states as functions of the junction position $x$. The position $x = 0$ represents the interface between the leftmost FI and the ESs, while the position $x = 3\xi$ represents the interface between the ESs and the rightmost SC. The black, red, and blue lines denote the probability densities $\rho$ for these three states with the energies $E/\Delta = 0$, 0.39, and 0.76, respectively. Because the maximal probability is in the range $0 < x < 3\xi$, the bound states are localized mainly in the ES region. By comparing the probability densities $\rho$ of the aforementioned three states, we find that the MBSs are slightly more localized than the fermionic ABSs.

While coupling another SC--ES--FI junction to the right-hand side of Fig.~\ref{fig1}(a), we create an FI--ES--SC--ES--FI junction, as shown in Fig.~\ref{fig1}(b). If the length $l_{\rm{SC}}$ of the SC is sufficiently large, each energy $E$ corresponds to two degenerate bound states, which are localized mainly at the left and right ES regions, respectively. If $l_{\rm{SC}}$ is not sufficiently large, the two degenerate bound states are coupled and then split into two non-degenerate states. As shown in Fig.~\ref{fig2}(c), the MBS with $E = 0$ splits into two states with $E = \pm0.02$, the fermionic ABS with $E = 0.39$ splits into two states with $E = 0.37$ and $0.41$, and the fermionic ABS with $E = 0.76$ splits into two states with $E = 0.74$ and $0.79$. Compared with the uncoupled states in Fig.~\ref{fig2}(b), we find that the maximum probability density $\rho$ is halved when the bound states are coupled in Fig.~\ref{fig2}(c). This sharp decrease in $\rho$ indicates that the degenerate bound states are well coupled. Furthermore, the amplitude of coupling is nearly same because the probability densities $\rho$ differ only slightly in Fig.~\ref{fig2}(c). In brief, both MBSs and fermionic ABSs show nonlocality in Fig.~\ref{fig2}(c). Herein, we consider $\mu_{\rm{FI}1} = \mu_{\rm{FI}2} = 0$, $\mu_{\rm{ES}1} = \mu_{\rm{ES}2} = 10\Delta$, and $m_{lz} = m_{rz} = \Delta$.

\begin{figure}[t]
\includegraphics[width=0.47\textwidth]{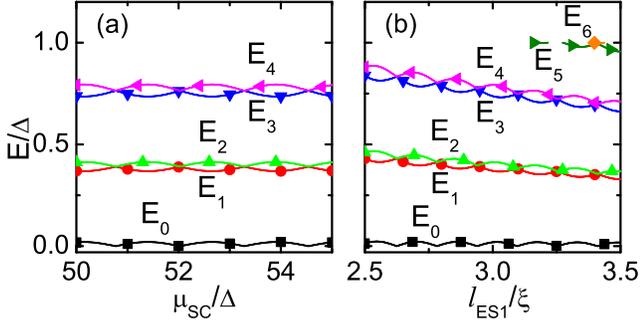}
\caption{(Color online) (a) and (b) Energies $E$ of the bound states as functions of the chemical potential $\mu_{\rm{SC}}$ of the SC and the width $l_{\rm{ES}1}$ of the leftmost ES in the FI--ES--SC--ES--FI junction. Here, the width $l_{\rm{SC}}$ of the SC is $2.5\xi$. (a) $l_{\rm{SC}} = 2.5\xi$, and $l_{\rm{ES}1} = l_{\rm{ES}2} = 3\xi$. (b) $\mu_{\rm{SC}} = 50\Delta$, and $l_{\rm{ES}1} = l_{\rm{ES}2}$.}
\label{fig3}
\end{figure}
Because the couplings of the MBSs and fermionic ABSs are considerably important, we extend the scope of the study to investigate the coupling properties of all pairs of degenerate bound states in Fig.~\ref{fig3}. To ensure the formation of twofold degenerate bound states, the lengths of the ESs on either side of the SC are configured to be the same in Fig.~\ref{fig1}(b). First, we plot the dependence of the energies of all bound states on the chemical potential $\mu_{\rm{SC}}$ in Fig.~\ref{fig3}(a). As $\mu_{\rm{SC}}$ is increased, we find that each energy pair $E$ periodically oscillates with a constant amplitude. Figure~\ref{fig3}(b) shows how the bound state energies depend on the width $l_{\rm{EG1}}$. We find that each energy pair oscillates with an increase in $l_{\rm{ES1}}$. Because the energies of the degenerate fermionic ABSs decrease with an increase in $l_{\rm{ES1}}$, as shown in Fig.~\ref{fig2}(a), the energy of each pair of the corresponding bound states decreases with an increase in $l_{\rm{ES1}}$ as a whole. Therefore, the coupling strength of each bound state pair decreases slightly and periodically with an increase in $l_{\rm{ES1}}$ overall in Fig.~\ref{fig3}(b).

In Fig.~\ref{fig3}, we see that the coupling strengths of all bound state pairs display the same tendency with the increase of $\mu_{\rm{SC}}$ or $l_{\rm{ES1}}$. Concretely, the coupling strengths maximize/minimize in phase with each other. As discussed below, this property, along with the nonlocality of MBSs and fermionic ABSs, is very important for the transport properties.

\emph{CCCs.---} We study the transport properties of the MBSs and ordinary fermionic ABSs in the junction shown in Fig.~\ref{fig1}(b). This can be realized by connecting the junction to two separate ES leads, whereupon the transport setup becomes the junction shown in Fig.~\ref{fig1}(c). By solving Eq.~(\ref{BdG Hamiltonian}), we can obtain the wave functions shown in Fig.~\ref{fig1}(c) and match them at the opposite sides of the six interfaces, namely at $x = -l_{\rm{FI1}}-l_{\rm{ES1}}$, $-l_{\rm{ES1}}$, $0$, $l_{\rm{SC}}$, $l_{\rm{SC}}+l_{\rm{ES2}}$, and $l_{\rm{SC}}+l_{\rm{ES2}}+l_{\rm{FI2}}$. The scattering matrix $S$ can then be obtained as follows:
\begin{equation}
\label{scattering matrix}
S=
  \begin{pmatrix}
    s^{ee}_{11} & s^{ee}_{12} & s^{eh}_{11} & s^{eh}_{12} \\
    s^{ee}_{21} & s^{ee}_{22} & s^{eh}_{21} & s^{eh}_{22} \\
    s^{he}_{11} & s^{he}_{12} & s^{hh}_{11} & s^{hh}_{12} \\
    s^{he}_{21} & s^{he}_{22} & s^{hh}_{21} & s^{hh}_{22} \\
  \end{pmatrix}.
\end{equation}
Based on the scattering matrix $S$, we can calculate the time-averaged current $\bar{I}_{i}$ and the current fluctuations $\delta I_{i}(t) = I_{i}(t)-\bar{I}_{i}$ in lead $i$. In our setup shown in Fig.~\ref{fig1}(c), the left and right leads are equally biased at voltage $V$, whereas the middle SC is grounded. The Fano factor measures the charge transfer in a current pulse, which is defined by the ratio of the noise correlator $P_{ij}$ to the mean current $\bar{I}_{i}$. The noise correlator $P_{ij}$ is defined as $P_{ij} = \int_{-\infty}^{\infty}dt\overline{\delta I_{i}(0)\delta I_{j}(t)}$. According to the scattering matrix elements in Eq.~(\ref{scattering matrix}), the mean current and noise correlator can be calculated as follows~\cite{M.P.Anantram1996}:
\begin{eqnarray}
\bar{I}_{i} = \frac{\mathrm{e}}{\mathrm{h}}\sum_{k\in1,2; \beta,\gamma\in e,h}\mathrm{sgn}(\beta)\int_{0}^{\infty}dE A_{kk}^{\gamma\gamma}(i,\beta,E)f_{i,\beta}(E),\nonumber
\end{eqnarray}
\begin{eqnarray}
A_{kl}^{\gamma\delta}(i,\beta,E) = \delta_{ik}\delta_{il}\delta_{\beta\gamma}\delta_{\beta\delta}-(s^{\beta\gamma}_{ik})^{\ast} s^{\beta\delta}_{il},
\end{eqnarray}
\begin{eqnarray}
P_{ij}& = &\frac{\mathrm{e}^{2}}{\mathrm{h}}\sum_{k,l\in1,2; \beta,\gamma,\zeta,\eta\in e,h}\mathrm{sgn}(\beta)\mathrm{sgn}(\gamma)\int_{0}^{\infty}dE A_{kl}^{\zeta\eta}(i,\beta,E)\nonumber\\
&&A_{lk}^{\eta\zeta}(j,\gamma,E)f_{i,\beta}(E)[1-f_{j,\gamma}(E)],\nonumber
\end{eqnarray}
where $i$, $j$, $k$, and $l$ denote the channels. For example, $k = 1$ and $2$ indicate the two channels in the left and right leads, respectively. The electron ($e$) and hole ($h$) channels are denoted by $\beta$, $\gamma$, $\zeta$, and $\eta$. Here, $\mathrm{sgn}(\beta) = 1$ for $\beta = e$ and $\mathrm{sgn}(\beta) = -1$ for $\beta = h$. The differential conductance in lead $i$ is $G_{i}=d \bar{I}_{i}/dV$, and $G_{1}$ is equal to $G_{2}$ because the bias voltage $V$ in the two leads is the same. The differential noise correlator is defined as $\mathcal{P}_{ij}(E) = d P_{ij}/d(eV)$. It is caused by electrons with energy $E$ and measures the CCC between the leads $i$ and $j$. To make the cross CCC sufficiently large, the left and right FIs are set to be adequately long to enhance the crossed Andreev reflection in Fig.~\ref{fig1}(c).

\begin{figure}[t]
\includegraphics[width=0.47\textwidth]{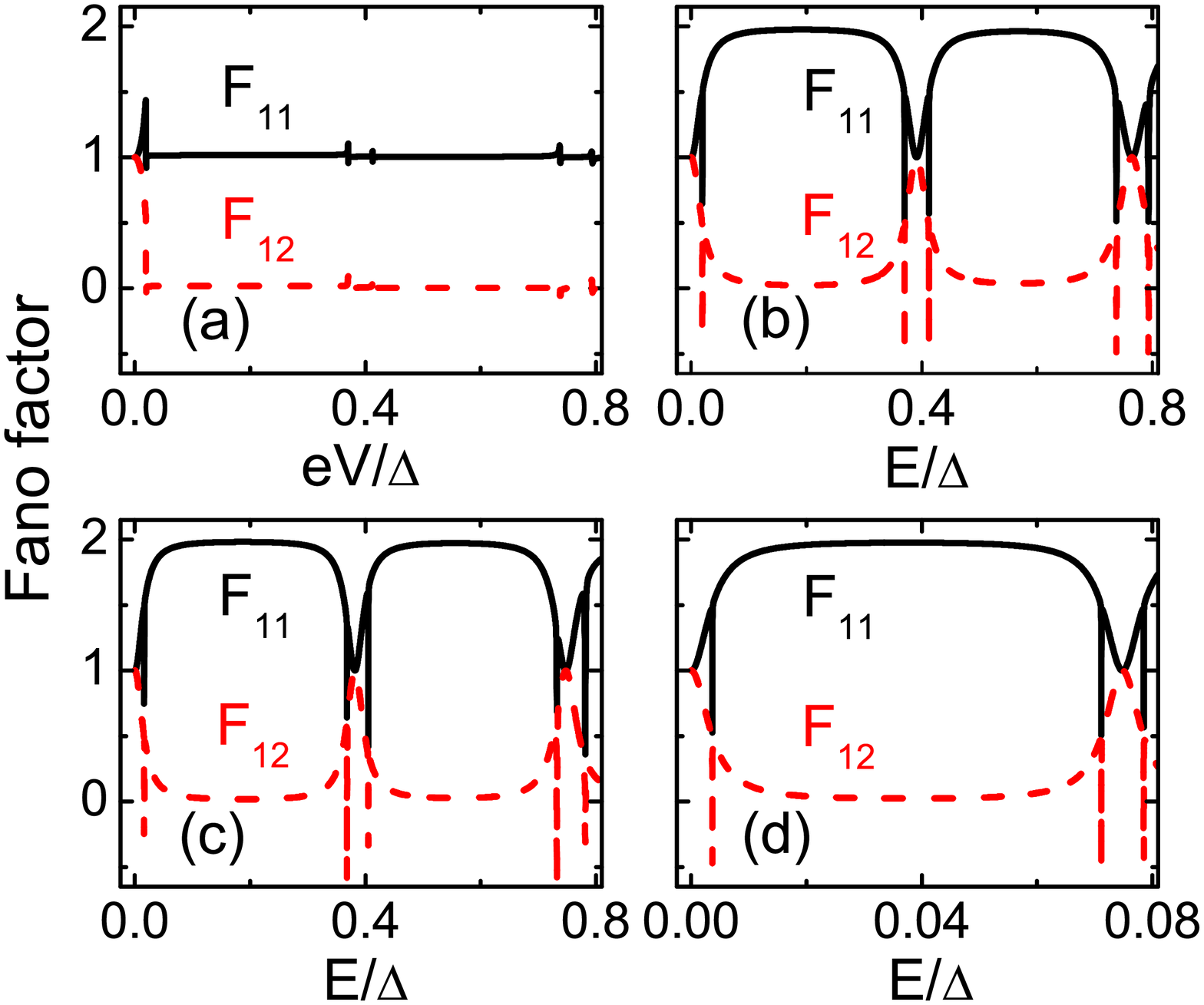}
\caption{(Color online) (a) Fano factors of the junction as functions of bias voltage $V$. (b)--(d) Fano factors as functions of energy $E$ of incident electrons. Here, the chemical potential is $\mu_{\rm{SC}} = 50\Delta$, $l_{\rm{SC}} = 2.5\xi$, and $l_{\rm{FI1}} = l_{\rm{FI2}} = 4\xi$. $l_{\rm{ES1}} = l_{\rm{ES2}} = 3\xi$ in (a) and (b), $l_{\rm{ES1}} = 3\xi$ and $l_{\rm{ES2}} = 3.1\xi$ in (c), and $l_{\rm{ES1}} = l_{\rm{ES2}} = 20\xi$ in (d).}
\label{fig4}
\end{figure}

Figure~\ref{fig4}(a) shows the current--current fluctuation correlators, which are calculated at zero temperature and represented by the Fano factors $F_{11}$ and $F_{12}$. $F_{11}$ denotes the autocorrelator $P_{11}$, which is normalized by $e\bar{I}_{1}$, and $F_{12}$ denotes the cross correlator $P_{12}$, which is normalized by $e\bar{I}_{1} = e\bar{I}_{2} = e(\bar{I}_{1}+\bar{I}_{2})/2$. Figure~\ref{fig4}(a) plots the dependence of $F_{11}$ and $F_{12}$ on the bias voltage $V$, and we observe that $F_{11}$ and $F_{12}$ are both equal to unity at $V = 0$. $F_{11} = 1$ indicates that the current pulse in lead $1$ transfers one electron into the SC, while $F_{12} = 1$ signifies both suppression of the local Andreev reflection and enhancement of the crossed Andreev reflection. As pointed out in previous research~\cite{J.Nilsson2008}, for any stochastic process the cross correlator is bound by the autocorrelator with $|P_{12}|\leq(P_{11}+P_{22})/2$. At $V = 0$, we have $P_{12} = (P_{11}+P_{22})/2 = P_{11}$ because $P_{11} = P_{22}$, making the cross correlator positive and maximally large for each current pulse. When $V$ is away from zero, the Fano factors $F_{11}$ and $F_{12}$ are mainly equal to $1$ and $0$, respectively. Such signals imply that the current fluctuations in the two leads are independent. Note that for a given bias voltage ($V$), the current correlators and the mean currents are calculated by summing over all contributions from $E = 0$ to $eV$. In Fig.~\ref{fig4}(a), $F_{11}$ and $F_{12}$ show four small peaks at either side of $eV = 0.39$ and $0.76$. These peaks signify some unusual transport properties of ordinary fermionic ABSs.

In Fig.~\ref{fig4}(b), we show the dependences of the differential Fano factors $F_{11}$ and $F_{12}$ on the energy $E$ of the incident electrons, where $F_{11}(E) = \mathcal{P}_{11}/G_{1}$ and $F_{12}(E) = \mathcal{P}_{12}/[(G_{1}+G_{2})/2] = \mathcal{P}_{12}/G_{1}$ because $G_{1} = G_{2}$. We can observe that $F_{11} = F_{12}$ at three energy points, namely $E = 0$, $0.39$, and $0.76$. As pointed out in the aforementioned section, $E = 0$ corresponds to the MBS energy, whereas $E = 0.39$ and $0.76$ correspond to the energies of the two different fermionic ABSs. Therefore, the fluctuations of the currents flowing from the two leads into the SC are maximally correlated at the energies of the MBSs and fermionic ABSs. This considerably differs from previous work~\cite{J.Liu2013} in which MBS signatures in CCCs were distinct from those of fermionic ABSs. When the energy $E$ is away from those of the three bound states, $F_{11}$ reaches $2$ and $F_{12}$ reaches $0$, thereby demonstrating that only local Andreev reflection occurs in those regions. On comparing Fig.~\ref{fig4}(a) and (b), we find that the Fano factors show the same characteristics near $V = 0$ and $E = 0$. This type of characteristics is attributed to the weak integral effect over the energy from $E = 0$ to $E = eV$ when the bias voltage is small. Therefore, the manner in which $F_{11}$ and $F_{12}$ depend on energy $E$ can well reveal the properties of the CCCs induced by the MBSs and fermionic ABSs.

Next, we study the influence of the structural asymmetry on the CCCs. Figure~\ref{fig4}(c) shows the dependences of the Fano factors on energy $E$, where $l_{\rm{ES1}} = 3\xi$ and $l_{\rm{ES2}} = 3.1\xi$. We find that the maximal correlated fluctuations of the currents also exist for the MBSs and fermionic ABSs. Such correlations always appear when the asymmetry between $l_{\rm{ES1}}$ and $l_{\rm{ES2}}$ does not strongly break the coupling of the two bound states. Furthermore, we consider the transport properties of a fermionic ABS when its energy is considerably close to the MBS energy. As shown in Fig.~\ref{fig2}(a), fermionic ABSs with considerably low energy will appear when $l_{\rm{ES1}}$ and $l_{\rm{ES2}}$ are sufficiently large. Figure~\ref{fig4}(d) plots the dependences of the Fano factors on energy $E$, where the energies of the fermionic ABSs are close to the MBS energy and the cross CCCs induced by the MBSs and fermionic ABSs are both maximal.

\emph{Conclusion.---}We studied the MBSs and ordinary fermionic ABSs in a 2D TI system. Our findings reveal that both MBSs and fermionic ABSs have nonlocality and that the coupling strengths of these bound states depend on the system parameters in the same pattern. When the integral effect on the bias voltage is eliminated, these two types of bound states can lead to the same differential CCCs. This characteristic demonstrates a universal property, and such CCCs are robust to the asymmetry of the system's structure.

This work was financially supported by the National Key Research and Development Program (Grant NoS.: 2017YFB0405703 and 2016YFA0301700), the National Natural Science Foundation of China (Grant Nos.: 11474265 and 11704366), China Postdoctoral
Science Foundation (Grant No.: 2016M590569), and the China Government Youth 1000-Plan Talent Program. We are grateful to the supercomputing service of AM-HPC and the Supercomputing Center of USTC for providing the high-performance computing resources.

\end{document}